\documentclass{PoS}
\usepackage{graphicx}

\newcommand {\be} {\begin{equation}}
\newcommand {\ba} {\begin{eqnarray}}
\newcommand {\ee} {\end{equation}}
\newcommand {\ea} {\end{eqnarray}}

\title{Electron-Positron Radiative
Annihilation : Timelike Virtual Compton Scattering}

\ShortTitle{Electron-Positron
Radiative Annihilation }

\author{\speaker{Asmita Mukherjee}%
        \thanks{In collaboration with A. Afanasev, S. J. Brodsky, C.
Carlson}\\
       Physics Department, IIT Bombay, Powai, Mumbai 400076 India\\
       E-mail: \email{mukherjee.asmita@gmail.com}}

\abstract{We report on a recent work proposing measurements of  
the deeply virtual Compton amplitude
(DVCS) $\gamma^* \to h \bar h \gamma$  in the timelike  
$ t = (p_{h} + p_{\bar h})^2 > 0$ kinematic domain which is accessible at 
electron-positron colliders via
the radiative annihilation process $e^+ e^- \to h \bar h \gamma.$
 }

\FullConference{Light Cone 2010 - LC2010\\
		June 14-18, 2010\\
		Valencia, Spain}

\begin{document}

\section{Introduction}

\begin{figure}[tbp]
\begin{center}

\vskip 2 mm

\includegraphics[width = 5.5 cm]{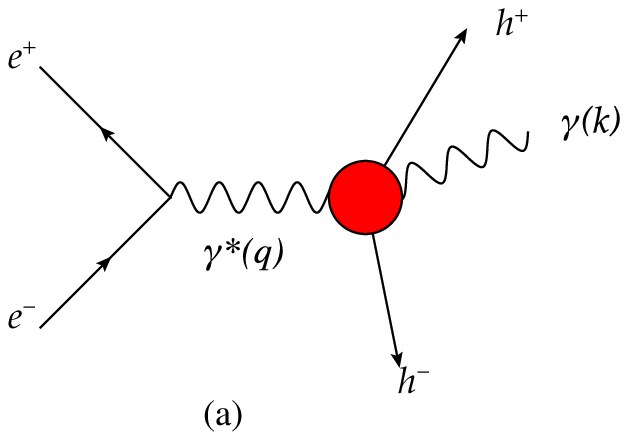}
\vskip 5 mm

\includegraphics[width = 8.4 cm]{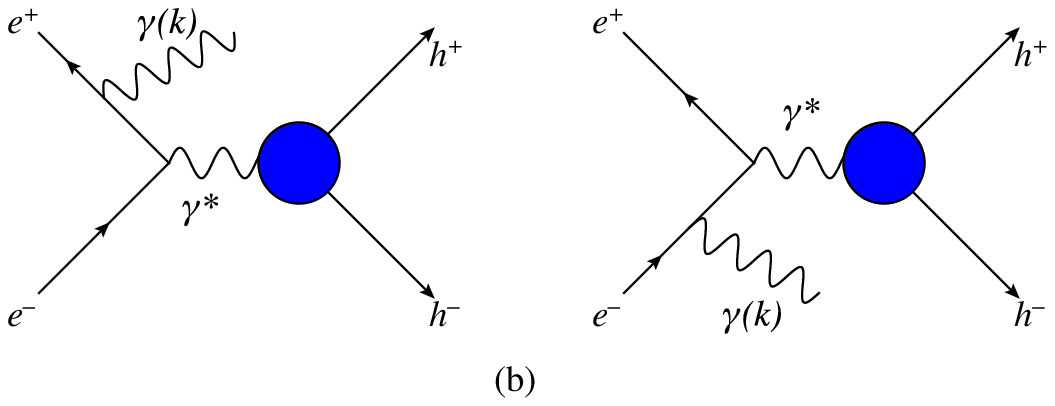}

\caption{Processes contributing to $e^+ e^- \to h^+ h^- \gamma$: (a)
the generic timelike DVCS process and (b) Bethe-Heitler processes.}
\label{fig:sjb1}
\end{center}
\end{figure}

Deeply virtual compton scattering (DVCS) process, $e p \to e^\prime \gamma
p^\prime,$ where the  intermediate photon  virtuality
$q^2 = (p_e^\prime - p_e)^2 < 0 $ and the momentum transfer to the
target proton $t = (p^\prime - p)^2 <0$  are spacelike;
 provides access to the generalized
parton distributions (GPDs)  of the  proton. They provide important information on the spin and spatial structure of the
nucleon. The GPDs measure hadron structure
at the amplitude level in contrast to the probabilistic properties
of  parton distribution functions. 
 In the forward limit (zero momentum transfer) they reduce to
ordinary parton distributions and $x$ moments of them are related to the
form factors.  The real part  of the DVCS amplitude is measured through the
interference with the Bethe-Heitler process and the imaginary part is
measured through various spin asymmetries. In contrast, the process we consider is the
radiative annihilation process $e^+ e^- \to h \bar h \gamma$~\cite{Lu:2006ut}, which
is accessible at electron-positron colliders and measures the timelike DVCS
amplitude ${\cal M}(\gamma^*(q) \to \gamma h  \bar h )$ illustrated in
Fig.~\ref{fig:sjb1}(a). We will discuss possible measurements of  the 
DVCS amplitude in the timelike or
$t >0$ kinematic domain, where $t = W^2 = (p_{h} + p_{\bar h} )^2$ is the
mass of the produced hadron pair. The hadronic matrix element is 
$C$-even since two photons attach to it.
The same final state can also come from Bethe-Heitler processes,
Fig.~\ref{fig:sjb1}(b), where the hadronic part of the matrix element
is $C$-odd. In addition to the process under consideration, doubly virtual 
Compton scattering, where one uses $ e^+ e^-
\to  e^+ e^- h \bar h $ to measure the amplitude ${\cal M}(\gamma^*(q)
\gamma^*(q') \to  h  \bar h )$ with one or both initial photons highly
spacelike, can also be measured at an  electron-positron collider.

One can apply charge conjugation to the electron and positron in the
initial state, thus relating two kinematic situations where the momentum
and spin of the electron and positron are interchanged.  The amplitudes
change sign or not depending on the photon attachment to the initial
electron line.  The asymmetry obtained by interchanging the electron and
positron is sensitive to the interference term between the $C$-even and
$C$-odd amplitudes as \cite{time}
\ba
\label{eq:asym}
A &=& \frac { \sigma - \sigma(e^+ \leftrightarrow e^- ) }{
\sigma +  \sigma(e^+ \leftrightarrow e^- )}
                \nonumber \\
&=&  \frac{ 2\, {\rm  Re}( {\cal M}^\dagger(C=+)  \times
{\cal M} (C= - ))  }
{ |{\cal M}(C=+) |^2 +  |{\cal M}(C= - ) |^2 }          \,,
\ea
which is sensitive to the relative phase of the $C$-even DVCS amplitude
and the timelike form factors.
The QED equivalents of these amplitudes,  where hadrons are replaced
by muons, usefully show that the magnitude of the $e^+ \leftrightarrow e^-$
asymmetry can be quite large.


We present a model calculation of the asymmetry for kinematic
conditions of existing electron-positron colliders.  Relevant
kinematics is chosen for tau-charm factories, s=14 GeV$^2$ (BEPCII) and
B-factories,  s=112 GeV$^2$ (Babar at PEPII and Belle at KEKB).

Measurements of the radiative annihilation process can provide valuable
new information on the analytic continuation of the DVCS amplitude,
including the existence of a $J=0$ fixed pole \cite{fixed},
which would be an amplitude that is constant in energy (and real in the
spacelike case) though not constant in momentum transfer.

We obtain a simple hadronic estimate by modeling the $C$-even $p \bar p$
timelike hadronic DVCS amplitude after an analysis of how it can be written
in terms of several Lorentz structures multiplied by $C=+$ form factors.
One of these is $R_V(\xi,W^2)$ and all of these can be related to the
timelike generalized parton distributions, or generalized distribution
amplitudes~\cite{Lansberg:2006fv,Diehl:2000uv}.  We will keep only
the $R_V$ term, which has the appearance of the QED amplitude multiplied by
$R_V(\xi,W^2)$, and we will model this form factor in a simplified way
where it depends only on $W^2 = t = (q-q')^2$ and is independent of $s$,
or the overall $q^2$.   One can say that this model simulates the C-even
Compton amplitude as a $J=0$ fixed pole amplitude with Regge behavior
$s^0$ at fixed $t$.
Of relevance here is an experimental result for the spacelike $C=+$ form
factor $R_V(t)$ from real 
wide-angle Compton scattering.  It is defined as the ratio of  the measured
real Compton amplitude 
$M(\gamma p \to \gamma^\prime p^\prime)$ divided by  the pointlike
Klein-Nishina formula.  $R_V(t)$ is 
measured to fall off as $1/ t^2$ at large $t$~\cite{Danagoulian:2007gs},
consistent with PQCD and 
AdS/QCD counting rules, which in turn is consistent with what we do in the
present context.

\section{Cross section and Asymmetry}
The process that we consider is, 
\be
\label{eq:tdvcs}
e^+(p_{e^+}) + e^-(p_{e^-}) \to p(p_{h^+}) + \bar p(p_{h^-}) + \gamma(q')
\ee
and for comparison, we also consider the same  process with $p$ and $\bar p$
replaced by $\mu^+$ and $\mu^-$, respectively.

The cross section for process~(\ref{eq:tdvcs}) is~\cite{Lu:2006ut}
\be
d\sigma = \frac{\beta W (s-W^2) }{64 (2\pi)^5 s^2 }  \left| \cal M \right|^2
dW d\Omega^*
d\Omega  \,,
\ee
where $| {\cal M} |^2$ is the matrix element summed over final and
averaged over initial polarizations and we also use the notations
\ba
s =q^2 = Q^2; ~~~~
s' = W^2 ;~~~~~
\beta = \sqrt{1 - \frac{4m^2}{W^2}}
\ea
The solid angle $\Omega^*$ gives the direction of the outgoing proton
or $\mu^+$ in the $p \bar p$ or $\mu^+ \mu^-$ rest frame and $\Omega$
gives the direction of the incoming electron in the $e^+ e^-$ rest frame.

\begin{figure}[tbp]
\begin{center}
   
\includegraphics[width = 8.4 cm]{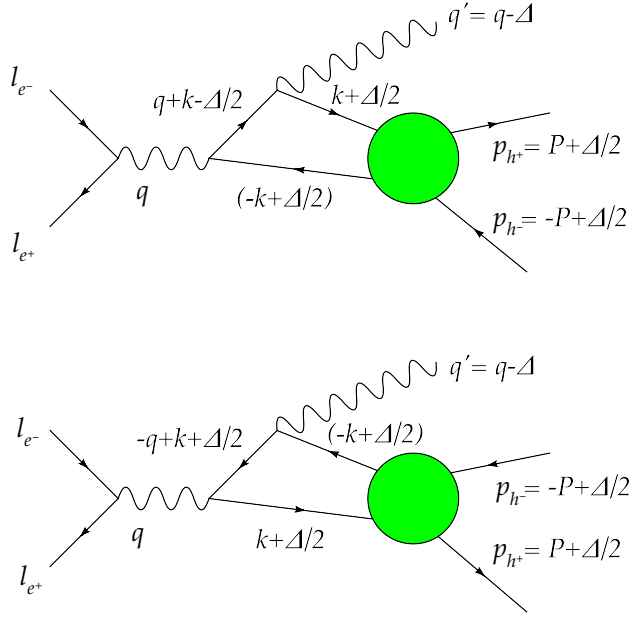}
\caption{Partonic diagrams for the case that the external photon is emitted
from the hadrons.}
\label{fig:tdvcs}
\end{center}
\end{figure}

Neglecting components that do not give large contributions in the Bjorken
limit, the amplitude becomes

\ba
{\mathcal M}^{\mu\nu} &=&  \frac{e_q^2}{2} \left( g^{\mu\nu} - p^\mu n^\nu -
n^\mu p^\nu \right)
                \nonumber \\&&
\quad \times           \int dx \, \left\{ \frac{1}{x+\xi+i\eta} +
\frac{1}{x-\xi-i\eta} \right\}
                \nonumber \\&&
\quad \times \bar u(p_h) \left[ \not\! n H^q
        + \frac{i}{2m} \sigma^{\alpha\beta} n_\alpha \Delta_\beta E^q
\right] v(p_{\bar h})
                \nonumber \\&&
-  \frac{ i e_q^2}{2}  \varepsilon^{\mu\nu\alpha\beta} p_\alpha n_\beta
        \int dx \, \left\{ \frac{1}{x+\xi+i\eta} - \frac{1}{x-\xi-i\eta}
\right\}
                \nonumber \\&&
\quad \times \bar u(p_h) \left[ \not\! n \gamma^5 \tilde H^q
        + \frac{n \cdot \Delta}{2m} \gamma^5 \tilde E^q  \right] v(p_{\bar
h})
                \nonumber \\[1.3ex]&&
 \equiv  - e_q^2  g^{\mu\nu}_\perp
                 \bar u(p_h) \left( \not\! n R^q_V
        + \frac{i}{2m} \sigma^{\alpha\beta} n_\alpha \Delta_\beta R^q_T
\right) v(p_{\bar h})
                \nonumber \\&&
+   i e_q^2  \varepsilon^{\mu\nu\alpha\beta} p_\alpha n_\beta
        \bar u(p_h) \left( \not\! n \gamma^5 R^q_A 
        + \frac{n \cdot \Delta}{2m} \gamma^5 R^q_P  \right) v(p_{\bar h})
        \,.
\ea

The form factors $R^q_V$, $R^q_T$, $R^q_A$, and $R^q_P$
are~\cite{Radyushkin:1998rt,Diehl:1998kh}
\ba
\label{eq:RV}
&&R^q_V(\xi,W^2) = \int dx \frac{x}{x^2-\xi^2 - i\eta} H^q(x,\xi,W^2)   ,
\nonumber \\
&&R^q_T(\xi,W^2) = \int dx \frac{x}{x^2-\xi^2 - i\eta}  E^q(x,\xi,W^2)   ,
\nonumber \\
&&R^q_A(\xi,W^2) = \int dx \frac{\xi}{x^2-\xi^2 - i\eta}  \tilde
H^q(x,\xi,W^2)    ,
\nonumber \\
&&R^q_P(\xi,W^2) = \int dx \frac{\xi}{x^2-\xi^2 - i\eta} \tilde E^q(x,\xi,W^2)
\ea
   
\noindent  The full amplitude will depend on
\be
R_{V}(\xi,t) = \sum e_q^2 R_{V}^q(\xi,t),
\ee
with similar equations for $V \to T,A,P$.
The asymmetries, Eq.~(\ref{eq:asym}), arise from interference between the
$C$-odd and $C$-even amplitudes.

\begin{figure}[tbp]
\begin{center}

\includegraphics[width = 5.5 cm]{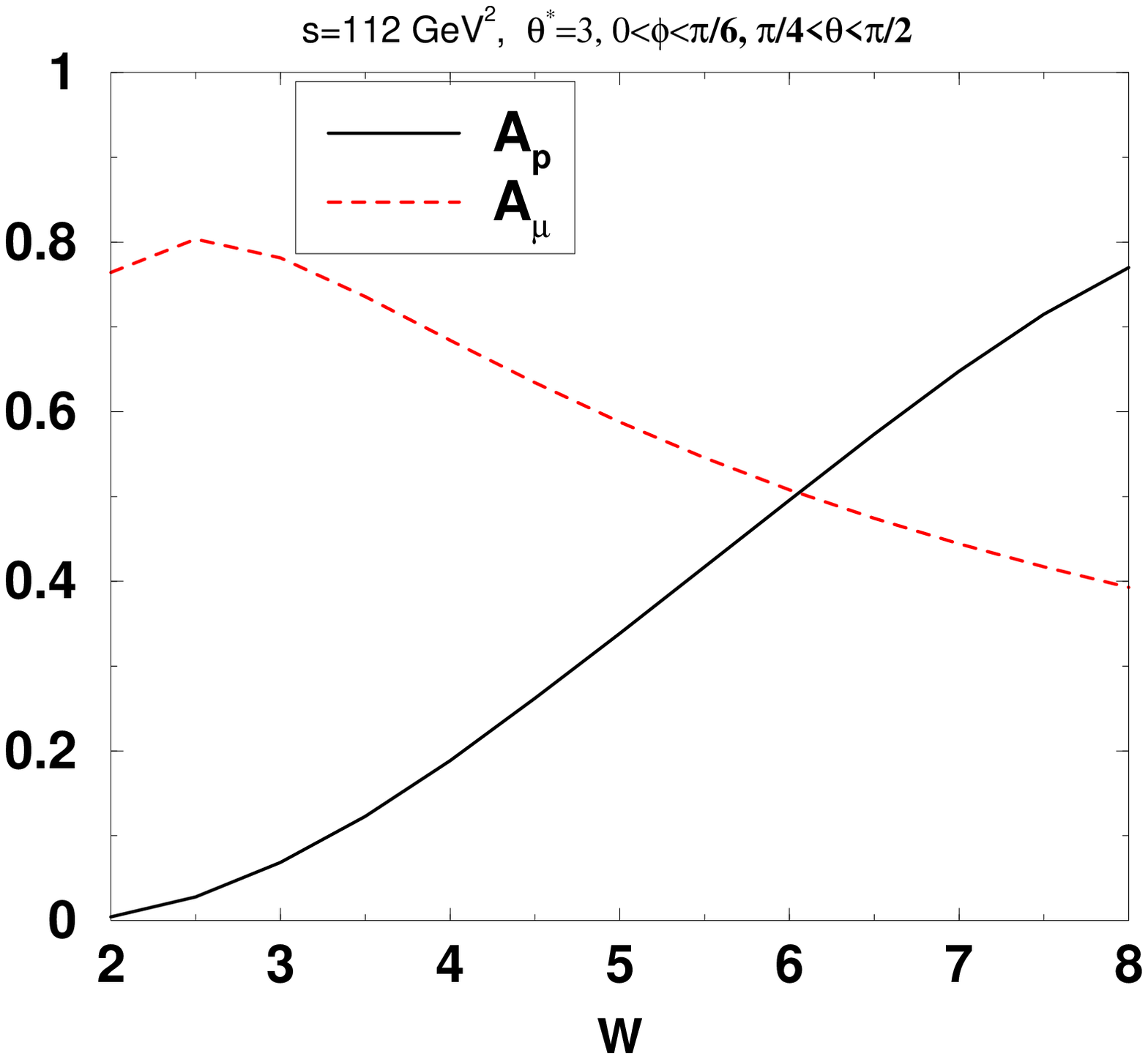}
\includegraphics[width = 5.5 cm]{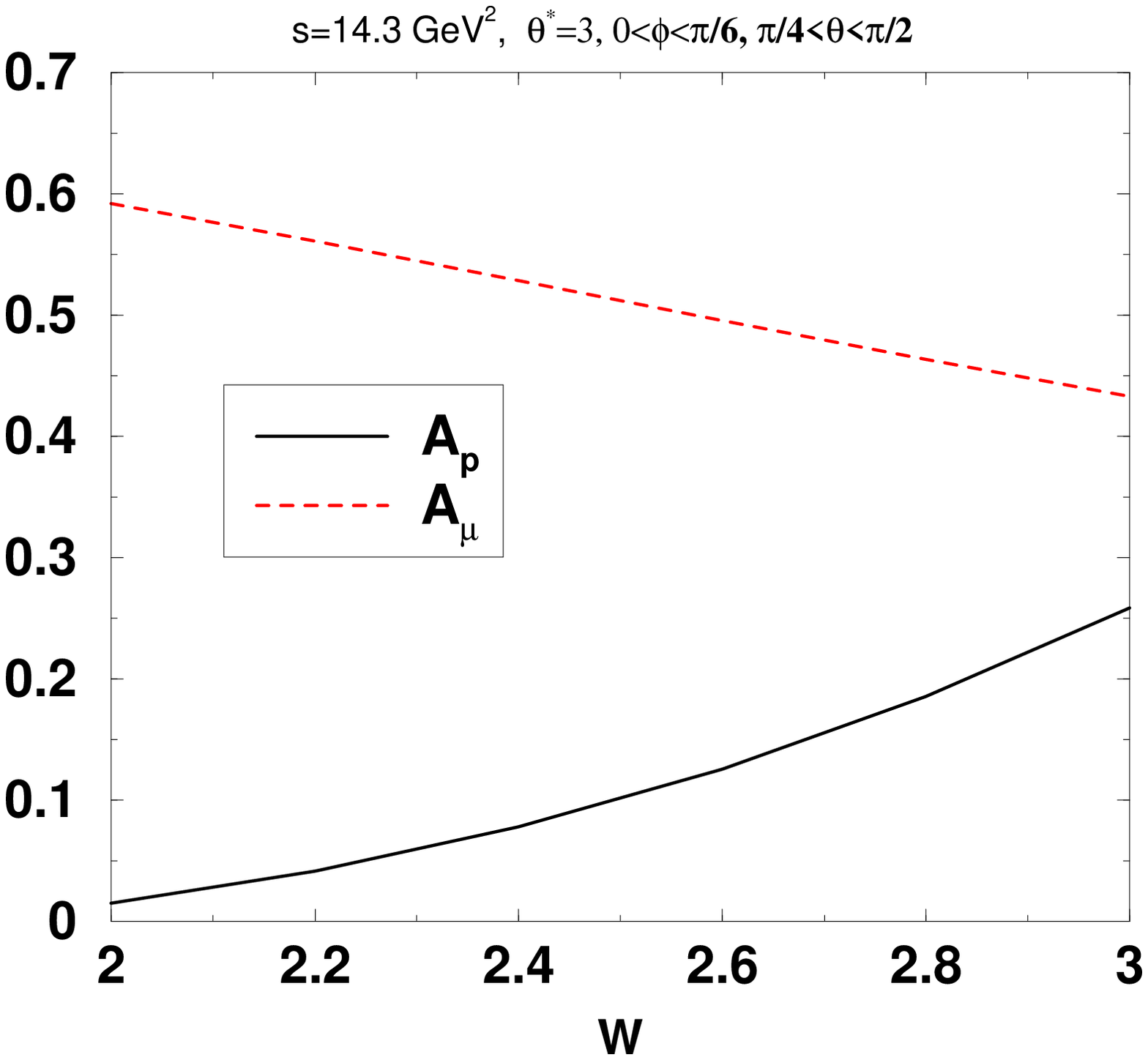}
\caption{Asymmetries for $\gamma^* \to \gamma p \bar p$ and its muonic
counterpart, plotted vs.
the final fermion pair invariant mass, over a range beginning close to the
$p \bar p$ threshold.
The upper graph ($s = 112$ GeV$^2$) is for BELLE or Babar energies, and the
lower graph ($s = 14.3$
GeV$^2$) is for BEPC II kinematics.  The angles (in radians) and angular
ranges are indicated on   
each plot. }
\label{fig:R_V}
\end{center}   
\end{figure}   

In modeling the hadronic part of the $C$-even diagrams, we keep just the
term $R_V(\xi,W^2)$.
Further, recall that photon-hadron amplitudes can expect a $J=0$ fixed pole,
which would be  a term with
flat energy dependence and a form factor like dependence on the momentum
transfer to the hadrons, $W^2$.
We give $R_V$ the same $W^2$ dependence as an electromagnetic form factor
and normalize by
\be
\label{eq:simpleapprox}
| R_V(\xi, W^2) | = \frac{4}{3} F_1(W^2)   \,.
\ee
The multiplicative factor is estimated from the expressions for
$R_V(\xi=0,W^2)$ and $F_1(W^2)$,
Eq.~(\ref{eq:RV}), and expecting domination by $u$-quarks
and approximate mean momentum fraction $x \approx 1/2$.

Support for this normalization and shape comes from data on spacelike wide
angle Compton scattering.
The numerically most important form factor here is $R_V(t)$, and data shows
that while $R_V(t)$ does drop
less rapidly with increasing $|t|$ than $F_1(t)$, it does not do so by a
lot, and that $R_V(t) = (4/3) F_1(t)$ is a decent representation of the data.

Further in our modeling, we note that the Lorentz structure that multiplies
$R_V(\xi,W^2)$ in the 
Bjorken limit is the same as one obtains from the QED amplitude in the
Bjorken limit.  If we are not 
deeply in the Bjorken region, we can argue that the Lorentz structure of the
amplitude is better 
represented by the QED amplitude, including the final fermion mass and
multiplied by $R_V(\xi,W^2)$.

Fig.~\ref{fig:R_V} shows 
two asymmetry plots, one at $s = 112$ GeV$^2$ relevant for Belle or Babar
energies and at 
$s = 14.3$ GeV$^2$ relevant for BEPC II energies.    The asymmetries are for
cross sections 
integrated over a stated range of angles, and plotted versus final hadronic
mass $W$.  
Since the sign of the asymmetry changes with $\phi$, one should not integrate
over more than half the 
range of that angle; if desired, one can integrate over fairly broad ranges
of $\theta$ and $\theta^*$.    
  For comparison, and to indicate the mass sensitivity for the selected $s$
and $W$, the plots also 
include the asymmetries expected for the purely muonic case.
The asymmetries can be large and measurable when the kinematics are 
well chosen. 

\section{Summary and Conclusions}
\label{sec:summary}

We studied deeply virtual Compton scattering amplitude in the timelike
domain in electron positron radiative annihilation process. We introduced a
forward-backward asymmetry to isolate the $C=+$ amplitude, where both
photons couple to the hadrons,  in an interference
with the $C=-$ amplitude where one one photon couple to the hadron. By
choosing a simple model we have showed that the asymmetry can be quite
large. 

\section{Acknowledgments}

We thank the organizers of LC2010 for the invitation and International Light
Cone Advisory Committee for Gary Mccartor Travel Award. 



\begin{thebibliography}{99}

\bibitem{time} A. Afanasev, S. J. Brodsky, C. E. Carlson, 
A. Mukherjee, Phys.Rev.{\bf D81}, 034014 (2010). 
e-Print: arXiv:0903.4188 [hep-ph]



\bibitem{Lu:2006ut}
  Z.~Lu and I.~Schmidt,
  Phys.\ Rev.\  D {\bf 73}, 094021 (2006)
  [Erratum-ibid.\  D {\bf 75}, 099902 (2007)]
  [arXiv:hep-ph/0603151].

\bibitem{fixed}
  S.~J.~Brodsky, F.~J.~Llanes-Estrada and A.~P.~Szczepaniak,
  Phys.\ Rev.\ D {\bf 79}, 033012 (2009)
  [arXiv:0812.0395 [hep-ph]].

\bibitem{Lansberg:2006fv}
  J.~P.~Lansberg, B.~Pire and L.~Szymanowski,
  Phys.\ Rev.\  D {\bf 73}, 074014 (2006)
  [arXiv:hep-ph/0602195].

\bibitem{Diehl:2000uv}
  M.~Diehl, T.~Gousset and B.~Pire,
  Phys.\ Rev.\  D {\bf 62}, 073014 (2000)
  [arXiv:hep-ph/0003233].

\bibitem{Danagoulian:2007gs}
  A.~Danagoulian {\it et al.}  [Hall A Collaboration],
  Phys.\ Rev.\ Lett.\  {\bf 98}, 152001 (2007)
  [arXiv:nucl-ex/0701068];
  A. Nathan, talk at the Workshop on Exclusive Reactions at High Momentum
Transfer
  (including results from JLab experiment E99-114),  May 21-24, 2007,
Jefferson Lab, Newport News, VA, USA

\bibitem{Radyushkin:1998rt}
  A.~V.~Radyushkin,
  Phys.\ Rev.\  D {\bf 58}, 114008 (1998)   
  [arXiv:hep-ph/9803316].

\bibitem{Diehl:1998kh}
  M.~Diehl, T.~Feldmann, R.~Jakob and P.~Kroll,
  Eur.\ Phys.\ J.\  C {\bf 8}, 409 (1999)
  [arXiv:hep-ph/9811253].


\end{thebibliography}
\end{document}